\documentclass[aps,prl,floatfix,twocolumn,preprintnumbers,amsmath,amssymb,groupedaddress,showpacs,showkeys]{revtex4-1}

\usepackage{graphicx}
\usepackage{amsmath}
\usepackage{amsfonts}
\usepackage{amssymb}
\usepackage{srcltx}
\usepackage{dcolumn}   
\usepackage{bm}        
\usepackage{color}     
\usepackage{psfrag}
\usepackage{booktabs}
\usepackage[breaklinks=true,colorlinks=true,linkcolor=blue,citecolor=blue,urlcolor=black,bookmarks=false]{hyperref}
\usepackage{todonotes}
\usepackage{graphicx}

\newcommand{\LIA}{\Lambda^{\rm{A}}_{\rm{I}}}
\newcommand{\LIB}{\Lambda^{\rm{B}}_{\rm{I}}}
\newcommand{\LBR}{\Lambda_{\rm{R}}}

\newcommand{\LPA}{\Lambda^{\rm{A}}_{\rm{PIA}}}
\newcommand{\LPB}{\Lambda^{\rm{B}}_{\rm{PIA}}}

\newcommand{\s}{\sigma}
\newcommand{\ii}{\mathrm{i}}

\newcommand{\T}{\mathrm{T}}

\def\nicefrac#1#2{
    \raise.1ex\hbox{#1}%
    \kern-.1em/\kern-.1em%
    \lower.3ex\hbox{#2}}

\begin{document}
\title{Theory of spin-orbit induced spin relaxation in functionalized graphene}

\author{Jan Bundesmann, Denis Kochan, Fedor Tkatschenko, Jaroslav Fabian, and Klaus Richter}
\affiliation{Institut f\"{u}r Theoretische Physik, Universit\"{a}t Regensburg, 93040 Regensburg, Germany}%

\begin{abstract}
We perform a comparative study of the spin relaxation by spin-orbit coupling induced from
adatoms (hydrogen and fluorine) in graphene. Two methods are applied, giving consistent results:
a full quantum transport simulation of a graphene nanoribbon, and a T-matrix calculation using Green's functions
for a single adatom in graphene. For hydrogenated graphene the dominant spin-orbit term for spin relaxation is PIA, the hitherto
neglected interaction due to pseudospin inversion asymmetry. In contrast, in fluorinated graphene PIA and Rashba
couplings destructively interfere, reducing the total spin relaxation rate. In this case we also predict a strong deviation
from the expected 2:1 spin relaxation anisotropy for out- and in-plane spin orientations. Our findings should be useful
to benchmark spin relaxation and weak localization experiments of functionalized graphene.
 \end{abstract}

\date{\today}
\maketitle

The spin properties of graphene derive, to a large extent, from what we combine it with \cite{Han2014:NatCom}.
In particular, functionalizing graphene with adatoms and molecules, which can induce giant spin-orbit couplings,
is a promising path towards practical graphene-based devices for spintronics \cite{Zutic2004:RMP, Fabian2007:APS}.
Indeed, adatoms such as hydrogen, fluorine, or copper, have been shown theoretically \cite{Neto2009:PRL, Gmitra2013:PRL, Irmer2015:PRB}
and experimentally, by measuring the spin Hall effect \cite{Balakrishnan2013:NP, Balakrishnan2014:NC}, to enhance the
spin-orbit coupling of graphene, from about 10~$\mu$eV in pristine graphene \cite{Gmitra2009:PRB}, to about 1-10~meV,
enough to cause sizable spin precession.

On the other hand, graphene's spin relaxation rate is also plagued by the (unintentional) extrinsic sources. The measured spin
lifetimes in graphene are on the orders of $0.1-1$~ns
\cite{Tombros2007:N,Tombros08, Pi2010:PRL, Yang2011:PRL,Han2011:PRL,Avsar2011, Jo11, Mani2012:NC,Guimaraes2014:PRL, Drogeler2014:NL,Kamalakar2015:NC},
which are consistent with the presence of magnetic moments on adatoms
or organic molecules \cite{Santos2012:NJP} chemisorbed on graphene, in both mono \cite{Kochan2014:PRL} and bilayer \cite{Kochan2015:ArX} forms. As little as 1 ppm of such moments can
induce ultrafast spin relaxation seen in spin injection experiments. The reason is that adatoms such as hydrogen, or organic
molecules, not only induce magnetic moments, but also act as resonant scatterers \cite{Wehling2010:PRL, Kochan2014:PRL, Bundesmann2014}.

Spin-orbit coupling (SOC) in graphene can still play an important role in three regimes:
(i) First, in the ultraclean high mobility limit in which SOC can induce a spin precession rate which is faster than the momentum relaxation rate. In such a case the averaging over the electron ensemble in the momentum space can lead to an ultrafast spin dephasing
\cite{Zutic2004:RMP}.
(ii) Second, in  graphene covered with heavy adatoms (Au, W, Tl) which typically sit on the hexagon centers and which induce strong SOC. The case of Au adatoms was recently investigated theoretically \cite{Tuan2014:NP}, discovering a new spin-pseudospin entanglement
mechanism of spin relaxation specific to graphene. The third regime, (iii), which is the subject of this paper, is graphene functionalized with light adatoms
(H, F, Cu) or molecules (CH$_3$) which prefer to sit at top (sometimes bridge) positions. While the induced SOC is much less than in (ii), such adatoms scatter Dirac electrons resonantly, strongly enhancing the spin-flip scattering and, with a sufficient coverage (0.01 - 1\%, depending on the adatom),
can overcome the dominance of local magnetic moments in spin relaxation and lead to sub 100~ps spin lifetimes.

We perform numerical and analytical investigations of the spin relaxation rates in graphene due to SOC of adatoms. We choose hydrogen and fluorine, for which realistic scattering potentials have been introduced from first-principles calculations \cite{Gmitra2013:PRL, Irmer2015:PRB}.
Hydrogen should be taken as an example of a class of adatoms with resonances close to the Dirac point, not literally as a hydrogen adatom, since hydrogen
also introduces local magnetic moments \cite{Yazyev2008:PRL} which would overshadow the effects of SOC in spin relaxation.
Fluorine, on the other hand, can be taken as a representative case for adatoms with resonances away the Dirac point, since
first-principles studies and experiments \cite{Hong2012:PRL} are inconclusive about its induced magnetic moment.
What we find is that the SOC resonantly enhances the spin relaxation rate. This is similar to the resonant enhancement
of the spin Hall effect due to adatoms, as recently predicted \cite{Ferreira2014:PRL,Pachoud2014:PRB}. Next, we show that the hitherto neglected
PIA (pseudospin inversion asymmetry) spin-orbit interaction is dominant in spin relaxation, relative to Rashba and intrinsic couplings, although in
fluorinated graphene PIA and Rashba destructively interfere. Finally, we find that while for hydrogen-like adatoms the
universal spin relaxation anisotropy of 2:1, for out- and in-plane spin relaxation, strictly holds, the anisotropy can be significantly
weaker for fluorine-like adatoms in which the intrinsic SOC becomes important at off-resonance energies.

\begin{figure*}
\includegraphics{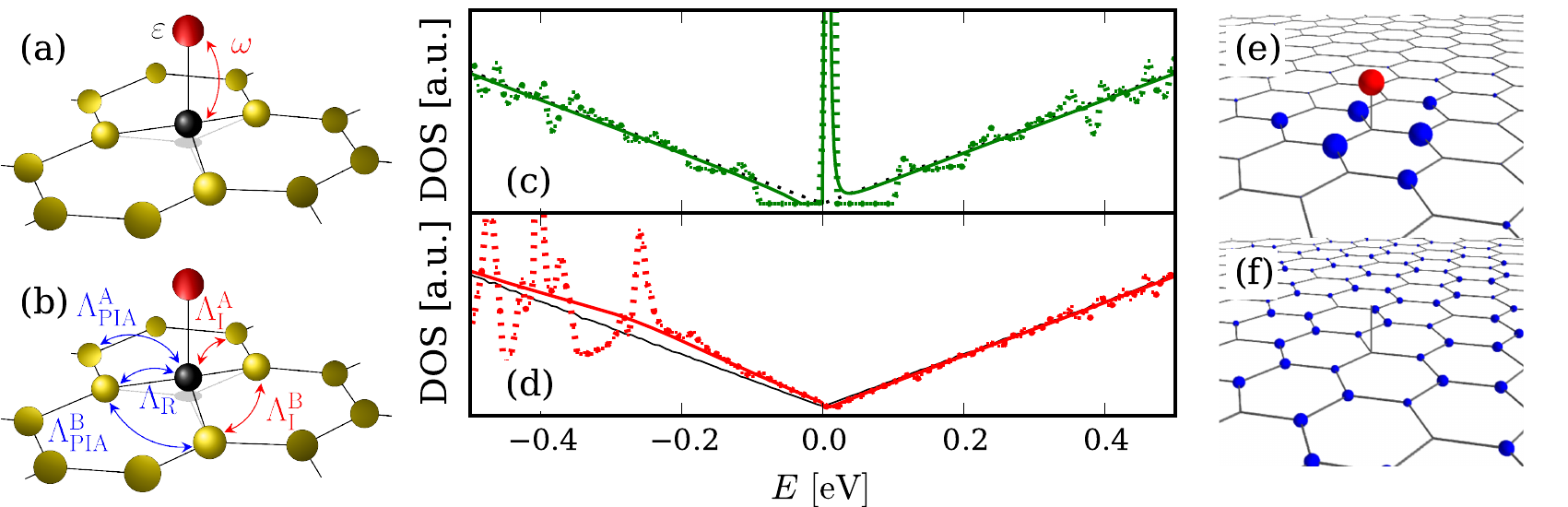}
\caption{(Color online) Adatoms on graphene. Panels (a) and (b) show the orbital and spin-orbit tight-binding hoppings, respectively, as described in text. The calculated DOS is in (c) for hydrogenated and in (d) for fluorinated graphene. Dotted lines are for a $20\times 20$ tight-binding supercell calculation (impurity concentration per carbon 0.125$\%$), while solid lines are obtained from the T-matrix single-adatom scattering formalism \cite{Irmer2015:PRB}. Pristine
graphene linear DOS is also given. Panels (e) and (f) compare the electron densities around a hydrogen adatom at
resonant ($E=7$\,meV) and off-resonant energies ($E=-500$\,meV).
}\label{Fig:hoppings_and_ldos}
\end{figure*}

We solve the spin relaxation problem due to adatom-induced SOC by two complementary techniques, a fully
numerical quantum transport simulation for a finite graphene stripe, and an analytical calculation of the T-matrix for
the given adatom. We bridge the two using the scattering formalism for graphene, showing that they give consistent results.
This by itself should be useful for other studies based on Landauer transport, for converting the transmission and reflection
probabilities to the scattering rates directly. We believe that our investigations of the spin relaxation due to SOC will serve as a
benchmark study for experimental investigations of the spin relaxation of functionalized graphene.

We focus on graphene with a dilute (such that scattering interferences can be ignored) coverage of adatoms that bond covalently to host's
carbons at top positions. Although our findings are general, we specifically consider hydrogen and fluorine adatoms for which we have realistic
Hamiltonians $\mathcal{H} = \mathcal{H}_{\rm gr} + V$, with $\mathcal{H}_{\rm gr}$
describing graphene (the usual nearest neighbor hopping Hamiltonian) and $V$ describing the adatom potential,
both orbital and spin-orbital. Parameters entering $V$ come from fits to first-principles data \cite{Gmitra2013:PRL, Irmer2015:PRB}.
These Hamiltonians are illustrated in Fig.~\ref{Fig:hoppings_and_ldos} and reproduced in detail in Suppl.~Material~(SM)~\cite{SM}. The orbital part,
in Fig.~\ref{Fig:hoppings_and_ldos}(a), comprises the on-site energy $\varepsilon$ and hopping $\omega$ between the adatom and the carbon atom
underneath. The spin-orbit part, in Fig.~\ref{Fig:hoppings_and_ldos}(b), has the intrinsic (I) spin-preserving intra-sublattice couplings $\LIA$ and $\LIB$,
the Rashba (R) inter-sublattice spin-flip hopping $\LBR$, and the pseudospin-inversion asymmetry (PIA)
terms $\LPA$ and $\LPB$ which couple the same sublattice with a spin flip \cite{Gmitra2013:PRL}.

Both hydrogen and fluorine are resonant adatoms, albeit with qualitatively different features.
The narrow hydrogen resonance \cite{Wehling2010:PRL,Kochan2014:PRL} is located close to the charge neutrality point at $E_{\mathrm{res}}\approx 7$\,meV with the full width at half maximum (FWHM) of $5$ meV; see the density of states (DOS) in Fig.~\ref{Fig:hoppings_and_ldos}(c).
The broad fluorine resonance \cite{Irmer2015:PRB} is centered at $E_{\mathrm{res}}\approx -260$\,meV with $\mathrm{FWHM}\approx 300$\,meV, see Figs.~\ref{Fig:hoppings_and_ldos}(d). To illustrate the resonant behavior we compare in Figs.~\ref{Fig:hoppings_and_ldos}(e) and (f) electron
densities around a hydrogen adatom at resonant and off-resonant energies. While off-resonance the states are delocalized, they are confined to the adatom
site at resonance. More details about local DOS analyzes are in SM \cite{SM}.

\begin{figure*}
\includegraphics[width=1.0\textwidth]{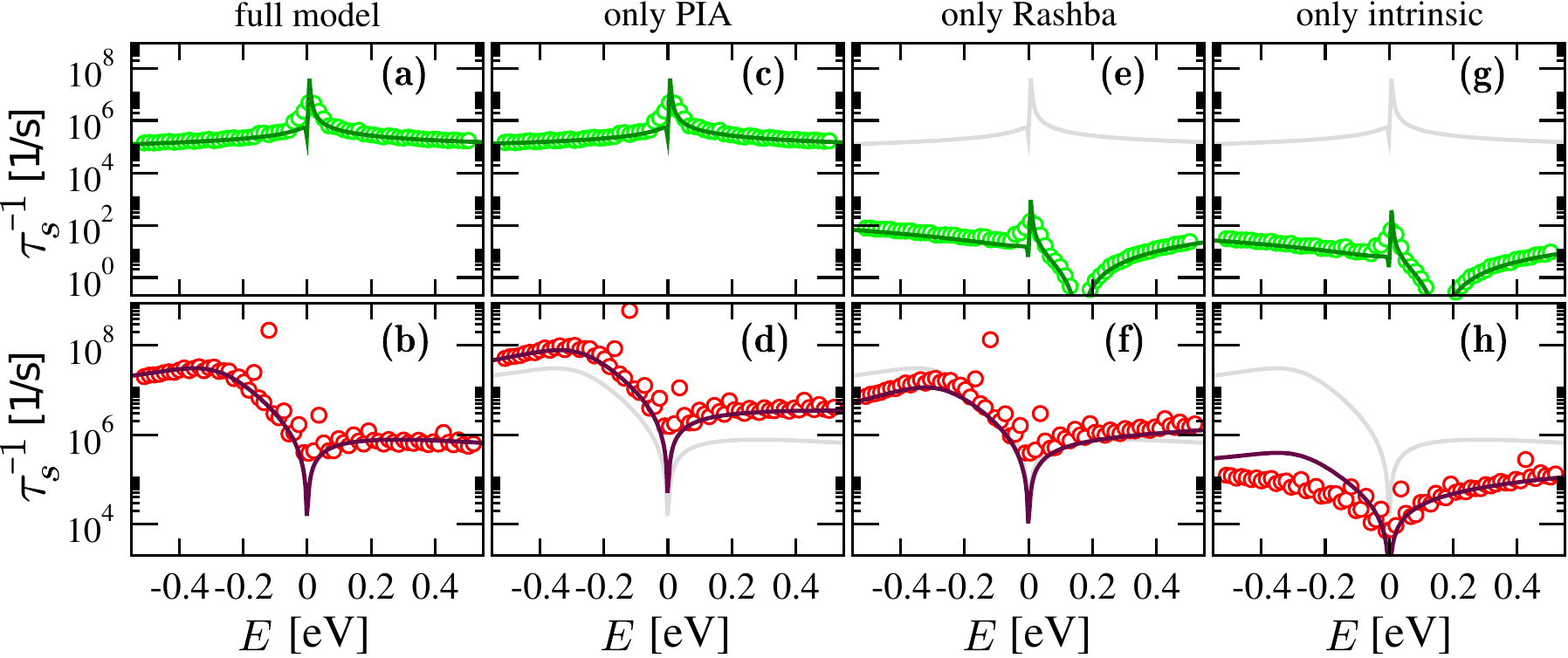}
\caption{(Color online) Calculated spin-relaxation rates as functions of the Fermi level for hydrogenated (upper panels) and fluorinated (lower) graphene, with impurity concentration $\eta=53$~ppm/carbon.
Symbols represent numerical Landauer-type calculations and solid lines come from analytical T-matrix analysis. Panels (a) and (b) show total spin-relaxation rates implementing all SOC terms.
Other columns show the spin-relaxation rates due to one SOC term only---PIA [panels (c), (d)], Rashba [panels (e), (f)], and intrinsic
[panels (g) and (h)] couplings; for comparison the total rates are reproduced as grey lines. The PIA coupling generally dominates over the Rashba and intrinsic ones, most pronounced for hydrogenated graphene.}
\label{Fig:rates}
\end{figure*}

We investigate spin transport and spin relaxation of Dirac electrons in the presence of adatoms by two
complementary approaches: (i) numerical calculation of a real-space Green's function for a graphene
nanoribbon \cite{Liu2012:PRB, Wimmer2009} of width $W$, with periodic boundary conditions along a transverse direction, and (ii) an analytical non-perturbative
calculation of spin scattering relaxation rates. Both techniques are well established and we summarize their application to our problem in
SM \cite{SM}. In (i) we obtain the spin-flip probability
\begin{align}
\Gamma_s(E) = \sum_{\sigma \in \lbrace \pm 1 \rbrace} \sum_{i, j} \bigl(\rvert t_{i,j; \sigma, -\sigma}\rvert^2 + \rvert r_{i,j; \sigma, -\sigma}\rvert^2\bigr)
\end{align}
which is a function of the electron energy $E$. In the above formula $t$ and $r$ are the transmission
and reflection amplitudes between left-right propagating modes $i$ and $j$ of opposite spins $\sigma$. In approach (ii) we get
directly the spin relaxation rate $1/\tau_s(E)$ from the T-matrix,
\begin{align}\label{Eq:T-matrix}
\T_{m,\s|n,-\s}(E)=\bigl\langle m,\s\bigl|V\bigl[1-G(E)V\bigr]^{-1}\bigl|n,-\s\bigr\rangle\,,
\end{align}
where $m$ and $n$ label atomic positions, and $G(E)$ is the energy dependent Green's function of graphene
which is known analytically close to the Dirac point \cite{Wang2006:PRB,Sherafati2011:PRB,Ducastelle2013:PRB},
see also \cite{SM}, so we can obtain $\T(E)$ fully non-perturbatively and study resonances.

To connect the two approaches we need a relation between $\Gamma_s(E)$ and $1/\tau_s(E)$. To this end we solve the
Lippmann-Schwinger equation $|\Psi_E^+\rangle=|\mathrm{in}_E\rangle+G(E)\T(E)|\mathrm{in}_E\rangle$. Here
$|\mathrm{in}_E\rangle=|\mathbf{q},\tau,\sigma\rangle$ is the incoming unperturbed low-energy Bloch state (normalized to the
graphene unit cell), labeled with valley index $\tau=\pm 1$, momentum $\mathbf{q}$ (measured from the Dirac point $\tau\mathbf{K}$),
spin $\s=\pm 1$, and positive energy $E=\hbar v_F|\mathbf{q}|>0$; negative energies can be treated analogously.
Fixing an atomic site $m$ with a position vector $\mathbf{R}$ that lies far away from the impurity region we get for the amplitude
$\langle m|\mathrm{out}_E\rangle$ of the outgoing wave $|\mathrm{out}_E\rangle=G(E)\T(E)|\mathrm{in}_E\rangle$:
\begin{equation}\label{Eq:amplitude}
\langle m|\mathrm{out}_E\rangle\simeq \frac{e^{\ii |\mathbf{R}||\mathbf{q}|}}{\sqrt{|\mathbf{R}|}}
\sum\limits_{\tau'\in\{\pm\tau\} \atop \s'\in\{\pm\s\}}
\,e^{\ii\tau'\mathbf{K}\cdot\mathbf{R}}\,f_{\s'\s}^{\tau'\tau}(\mathbf{q}',\mathbf{q},E)\,,
\end{equation}
where $\mathbf{q}'=|\mathbf{q}|\tfrac{\mathbf{R}}{|\mathbf{R}|}$, $\mathbf{K}=\tfrac{4\pi}{3a}(1,0)$ ($a=2.46$\,{\AA} is the lattice constant),
and the partial amplitudes
\begin{equation}
f_{\s'\s}^{\tau'\tau}(\mathbf{q}',\mathbf{q},E)=-\sqrt{\frac{\ii E a}{\sqrt{12}\pi t^3}}\
\bigl\langle\mathbf{q}',\tau',\s'\bigr|\T(E)\bigl|\mathbf{q},\tau,\s\bigr\rangle\,.\label{Eq:scattering_amplitude}
\end{equation}
This matrix element can be obtained analytically---in the local atomic basis the T-matrix is a finite $20\times 20$-matrix,
see SM \cite{SM}.
Terms entering the sum in Eq.~(\ref{Eq:amplitude}) with $\tau'=\tau$ ($\tau'=-\tau$) contribute to the intra (inter) valley scattering,
while those with $\sigma'=\sigma$ ($\sigma'=-\sigma$) contribute to the spin conserving (spin flip) scattering.
The total spin-flip scattering cross-section for the energy $E$ and valley $\tau$ averaged over the incident directions then becomes
\begin{equation}\label{Eq:sigmaS}
\sigma_s^\tau(E)=\hspace{-2mm}\sum\limits_{\tau'\in\{\pm\tau\}\atop\sigma\phantom{'}\in\{\pm 1\}}\int\limits_{0}^{2\pi} \frac{\mathrm{d}\varphi_{_{\mathbf{q}}}}{2\pi}\int\limits_0^{2\pi}\mathrm{d}\varphi_{_{\mathbf{q}'}}
\bigl|f_{-\s\s}^{\tau'\tau}(\mathbf{q}',\mathbf{q},E)\bigr|^2\,,
\end{equation}
where $\varphi_{{\mathbf{q}^{(\prime)}}}$ is the polar angle of $\mathbf{q}^{{{(\prime)}}}$ with respect to $x$-axis.
For a nanoribbon with a width $W\gg\sigma_s^\tau$ the spin-flip scattering probability for the left-right charge-carriers transport equals
$\Gamma_s(E)=(1/2W)\sum_{\tau\in\{\pm\}}\sigma_s^{\tau}(E)$.
Forming the rate equations in terms of the T-matrix and summing over all processes at given energy $E$ we get for the spin-relaxation rate
$\tau_s^{-1}(E)$ at zero temperature
\begin{equation}\label{Eq:rateS}
\tau_s^{-1}(E)=\frac{2\pi}{\hbar}\,\eta\,\frac{t}{\pi a}\sum\limits_{\tau\in\{\pm\}}\sigma_s^\tau(E)=\frac{4t}{\hbar}\,\eta\,\frac{W}{a}\,\Gamma_s(E)\,.
\end{equation}
Here $\eta$ stands for the adatom concentration per carbon and $t=2.6$\,eV for the plane graphene hopping.

The main result of this paper is shown in Fig.~\ref{Fig:rates}. The Landauer transport data are obtained from zig-zag nanoribbon
of length $L=62a$ and width $W=130a$, with a single adatom impurity. This corresponds to $\eta=53$~ppm/carbon. The initial spin polarization is in the graphene plane. The agreement between the Landauer (i)
and the T-matrix (ii) approaches is remarkable, proving the validity of Eq.~(\ref{Eq:rateS}).

We first look at hydrogenated graphene, see Fig.~\ref{Fig:rates}(a). {\it The spin relaxation rate is strongly enhanced at resonance},
which is close to the Dirac point, following the analogous resonance dependence as the DOS in Fig.~\ref{Fig:hoppings_and_ldos}. At resonance the spin relaxation rate is an order of magnitude greater than off resonance. The shortest spin relaxation time at resonance is about 100~ns.
In the case of magnetic moments, the shortest spin relaxation time would be about 1~ps for this adatom concentration \cite{Kochan2014:PRL},
reaching the momentum relaxation time. The spin-orbit coupling mechanism is much less effective here, since the spin-orbit energy
is smaller than the resonance width. Typical the electron spends less time on the adatom than what would be required for
a spin precession by the adatom-induced SOC. For exchange coupling the situation is the opposite: during the
dwell on the adatom the spin can fully precess about the exchange field, so the spin flip and spin conserving scatterings are
equally likely. Recently it was demonstrated using first-principles quantum transport calculations \cite{Evers2015:ArX} that
the exchange coupling due to hydrogen adatoms in graphene nanoribbons can cause spin-flip conductance as large as the spin-conserving one,
confirming the resonance model picture.

Which of the three spin-orbit coupling terms, PIA, Rashba, and intrinsic, contribute most to the spin relaxation? We performed
calculations with the individual terms only and find that {\it PIA only is responsible for the spin relaxation in hydrogenated graphene
due to spin-orbit coupling}. Remarkably, the contributions from Rashba and intrinsic couplings are smaller by several orders of magnitude.
To explain this we analyzed the local DOS around the adatom site. At resonance, the electron density on carbon beneath the adatom
gets strongly reduced, see Fig.~\ref{Fig:hoppings_and_ldos}(e). Having two carbon sites $\mathrm{\mathrm{C}_1}$ and $\mathrm{\mathrm{C}_2}$ connected
by SOC hopping $\Lambda$, the effective spin-flip probability is directly proportional to $|\Lambda|^2\nu(\mathrm{\mathrm{C}_1})\nu(\mathrm{\mathrm{C}_2})$,
where $\nu(\mathrm{\mathrm{C}_1})$ and $\nu(\mathrm{\mathrm{C}_2})$ stand for the local DOS at those carbons. For the Rashba $\LBR$ and
intrinsic $\LIA$ SOC hoppings the affected carbon site is directly involved and hence we expect a weaker spin relaxation as compared
with $\LPB$ that connects carbon atoms on the populated sublattice; for hydrogen $\LIB$ and $\LPA$ vanish, see SM \cite{SM}.

We now turn to fluorinated graphene. The spin-orbit terms induced by fluorine are greater by decade compared to hydrogen, so
one expects the spin relaxation rate up to two decades faster. This is indeed what we find, as shown in Fig.~\ref{Fig:rates}(b).
However, the dependence of $1/\tau_s$ on energy looks very different from the hydrogen case. The rate has a broad peak
at negative energies, vanishes at the Dirac point, and becomes rather flat at positive energies. The broad feature at negative
energies can be connected to the resonance seen in the DOS in Fig.~\ref{Fig:hoppings_and_ldos}(d), implying resonance enhancement of the
spin relaxation. Close to the Dirac point, the DOS vanishes and the scattering theory predicts a vanishing scattering probability as well (unless the resonance
is close to that, as for hydrogen adatoms); this is why the spin relaxation rate has a dip there.

Resolving different spin-orbit terms uncovers a surprising effect of {\it destructive interference between the Rashba and PIA couplings}. Indeed,
considered individually, the spin relaxation rate due to the PIA interaction, as in hydrogenated graphene, is greater than the rate due to all interactions together, see Fig.~\ref{Fig:rates}.
The intrinsic term plays a minor role only, becoming important only at high positive energies. The destructive interference between PIA and Rashba SOC in the case of fluorinated graphene reduces the effective value of PIA, and with it the spin relaxation rate.
We identify this destructive path as the successive Rashba (spin-flip nearest neighbor) and orbital $t$ (spin-conserving nearest neighbor) hopping, which effectively leads to a next-nearest-neighbor spin-flip, just like PIA, but with the opposite sign. Details
are found in SM \cite{SM}.

\begin{figure}
\includegraphics[width=0.4\textwidth]{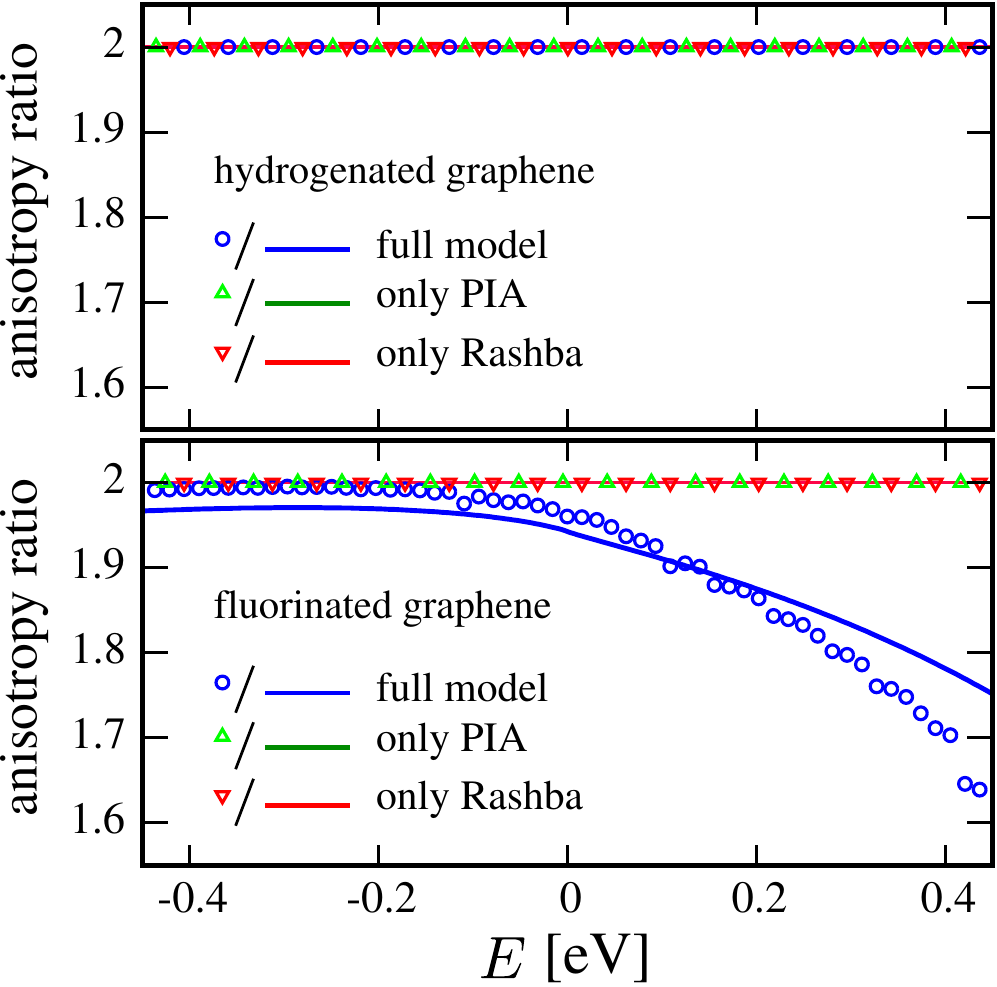}
\caption{(Color online) Calculated spin relaxation anisotropy as the defined ratio of $1/\tau_s$ for electron spins out- and in-plane, for hydrogenated (top panel) and fluorinated (bottom panel) graphene.
For the hydrogenated graphene the anisotropy ratio is 2 (all lines are on top of each other), as expected for spin-orbit fields; intrinsic coupling plays no role.
For the fluorinated graphene the anisotropy drops well below 2 at positive energies, as here also the intrinsic coupling becomes important.
Symbols represent numerical calculation and solid lines come from analytical model.}\label{Fig:anisotropy}
\end{figure}

Finally, we consider spin relaxation anisotropy, which is an experimentally important fingerprint of the SOC mechanism. Indeed, spin-orbit
fields (PIA and Rashba) lie in the graphene plane. An electron spin which points out of plane can be flipped by two independent components of a given
spin-orbit field, while an electron spin lying in the plane can be flipped by only one component (perpendicular to that spin). This gives the expected
2:1 ratio of the spin relaxation for out- and in-plane spins. This is also expected for graphene, if its spin relaxation is due to SOC.
For the hydrogenated graphene case our calculations show no deviation from this expectation, see Fig.~\ref{Fig:anisotropy}.
The intrinsic coupling contribution is weak, and the relaxation is
due to the spin-orbit fields only. In contrast, our calculations for fluorinated graphene show marked deviations from the 2:1 expectation at positive energies. At those energies the intrinsic SOC becomes also important, deforming the spin-flip picture due to spin-orbit fields only.

In conclusion, we showed that spin-orbit induced spin-relaxation in graphene functionalized with adatoms (we gave examples of hydrogen and fluorine) can exhibit a giant enhancement at resonances and,
for a sufficient adatom concentration, overcome the magnetic-moment limited spin relaxation. For both hydrogen and fluorine the PIA interaction gives the dominant contribution, although in fluorinated
graphene it interferes destructively with the Rashba coupling. Intrinsic SOC is inhibited in the spin relaxation processes, but can become important off resonance and even strongly modify the spin
relaxation anisotropy, providing an important signature that could be tested experimentally.

This work was supported by DFG~SFB~689, GRK~1570, Hans-B\"{o}ckler-Stiftung, and by the EU~Seventh~Framework~Programme under Grant~Agreement~No.~604391~Graphene~Flagship.

\bibliography{grp_adatom(2)}
\end{document}